\documentstyle[12pt,epsfig]{article}
\newcommand{\institute}[1]{\parbox{16cm}{%
\centering\normalsize \sl #1}}

\title{{\vspace{-0cm} \normalsize
\hfill \parbox[t]{4cm}{CERN-TH/96-57 \\ ITP-Budapest 517\\
                   }\\[7em]}
\bf The SU(2)-Higgs model on asymmetric lattices}
\author{%
F.~Csikor\\[0.5ex]
\institute{Institute for Theoretical Physics, E\"otv\"os University,\\
H-1088 Budapest, Hungary}\\[2ex]
Z.~Fodor%
\thanks{On leave from Institute for Theoretical Physics,
E\"otv\"os University, H-1088 Budapest, Hungary}\\[0.5ex]
\institute{Theory Division, CERN,\\CH-1211 Geneva 23, Switzerland}}
\date{}
\begin{document}
\maketitle
\begin{abstract} \noindent
We calculate the ${\cal O}(g^2,\lambda)$ corrections to the
coupling anisotropies of the SU(2)-Higgs model on lattices
with asymmetric lattice spacings.  
These corrections are obtained by a one-loop calculation
requiring the rotational invariance of the
gauge- and Higgs-boson propagators in the continuum limit.
\end{abstract}
\vfill
CERN-TH/96-57\\
February, 1996
\newpage
\section{Introduction}

At high temperatures the electroweak symmetry is restored.
Since the baryon violating processes are unsuppressed at high temperatures,
the observed baryon asymmetry of the universe has finally been
determined at the electroweak phase transition \cite{kuzmin}.

In recent years quantitative studies of the electroweak phase
transition have been carried out by means of resummed
perturbation theory
\cite{arnold}--\cite{buchmuller} and lattice Monte Carlo simulations
\cite{bunk}--\cite{jansen}. 
In the SU(2)-Higgs model for Higgs masses ($M_H$) below 50 GeV,
the phase transition is predicted by the perturbation theory to be 
of first order. However, no definite statement
can be made for physically more interesting masses, e.g.
$M_H>80$ GeV. Due to the bad infrared features of the theory, 
the perturbative approach breaks down in
this parameter region. A systematic and fully
controllable treatment is necessary, which can be achieved by
lattice simulations.

For smaller Higgs boson masses ($M_H<50$ GeV) 
the phase  transition is quite
strong and relatively easy to study on the lattice. 
For larger $M_H$ (e.g. $M_H=80$ GeV)
the phase transition gets weaker, the lowest excitations have
masses small compared to the temperature, $T$. From this
feature one expects that a finite temperature simulation
on isotropic lattice would
need several hundred lattice points in the spatial directions
even for $L_t=2$ temporal extension. These kinds of lattice sizes
are out of the scope of the present numerical resources.

One
possibility to solve the problem of these different scales is to
integrate out the heavy, ${\cal O}(T)$ modes perturbatively,
and analyse the obtained theory on the lattice. This strategy 
turned out to be quite successful, and both its perturbative
and lattice features have been studied by 
several groups \cite{kajantie}--\cite{philipsen}.

With this paper we follow another approach (analytic
and in the future Monte Carlo) to handle this 
two-scale problem. We will use the simple idea that
finite temperature field theory can be
conveniently studied on asymmetric lattices, i.e. lattices
with different spacings in temporal ($a_t$) and spatial
($a_s$) directions.  This method solves the two-scale 
problem in a natural way \cite{bender}. Another advantage is,
well-known and often used in QCD, that this formulation
makes an independent variation of the temperature ($T$) and
volume ($V$) possible. The perturbative corrections
to the coupling anisotropies are known in QCD (see 
refs. \cite{karsch82,karsch89}).

The plan of this letter is as follows. In Section 2 we give the lattice
action of the model on asymmetric lattices and discuss 
the effective potential.
Section 3 contains the calculation of the  
wave function quantum correction terms, which give 
the quantum corrections to the anisotropy 
parameters. Section 4 is devoted to the discussions and outlook.

\section{Lattice action and the critical hopping parameter}

For simplicity, we use equal lattice spacings in the three 
spatial directions ($a_i=a_s,\ i=1,2,3$) and another spacing
in the temporal direction ($a_4=a_t$). The asymmetry of the 
lattice spacings is characterized by the asymmetry factor $\xi=a_s/a_t$.
The different lattice spacings can be ensured by
different coupling strengths in the action for time-like and space-like
directions. The action reads
\begin{eqnarray}\label{lattice_action}
S[U,\varphi] &=& \beta_s \sum_{sp}
\left( 1 - {1 \over 2} {\rm Tr\,} U_{pl} \right)
+\beta_t \sum_{tp}
\left( 1 - {1 \over 2} {\rm Tr\,} U_{pl} \right)
\nonumber \\
&&+ \sum_x \left\{ {1 \over 2}{\rm Tr\,}(\varphi_x^+\varphi_x)+
\lambda \left[ {1 \over 2}{\rm Tr\,}(\varphi_x^+\varphi_x) - 1 \right]^2
\right. \nonumber \\
&&\left.
-\kappa_s\sum_{\mu=1}^3
{\rm Tr\,}(\varphi^+_{x+\hat{\mu}}U_{x,\mu}\,\varphi_x)
-\kappa_t {\rm Tr\,}(\varphi^+_{x+\hat{4}}U_{x,4}\,\varphi_x)\right\},
\end{eqnarray}
where $U_{x,\mu}$ denotes the SU(2) gauge link variable,  $U_{sp}$ and
$U_{tp}$
the path-ordered product of the four $U_{x,\mu}$ around a
space-space or space-time plaquette, respectively.
The symbol $\varphi_x$ stands for the Higgs field, which is
also written as $\varphi_x=\rho_x\cdot\alpha_x$, with $\rho_x \in
{\bf R^+}$ and $\alpha_x \in \mbox{\rm SU(2)}$.

The anisotropies
\begin{equation}
\gamma_\beta^2={\beta_t \over \beta_s}\ \ \ , \ \ \ \
\gamma_\kappa^2={\kappa_t \over \kappa_s}
\end{equation}
are functions of the asymmetry $\xi$. 
On the tree-level the coupling anisotropies are equal to the
lattice spacing asymmetry; however, they receive quantum corrections
in higher orders of the loop-expansion
\begin{equation}\label{tree}
\gamma_\beta^2=
\xi^2\left[1+c_\beta(\xi)g^2+b_\beta(\xi)\lambda
+{\cal O}(g^4,\lambda^2)\right],\ \ 
\gamma_\kappa^2=
\xi^2\left[1+c_\kappa(\xi)g^2+b_\kappa(\xi)\lambda
+{\cal O}(g^4,\lambda^2)\right].
\end{equation}
Here $g$ is the bare gauge coupling in standard notation and a 
formal double expansion in $g^2$ and $\lambda$ has been performed.
In this double expansion we use the formal power counting
$\lambda \sim g^2$.
It is useful to introduce the hopping parameter 
$\kappa^2=\kappa_s\kappa_t$ and
$\beta^2=\beta_s\beta_t$.
In general, the determination of $\gamma_\beta(\xi)$ and
$\gamma_\kappa(\xi)$
should be done non-perturbatively. This can be achieved by requiring
that the Higgs- and W-boson correlation lengths in physical units 
are the same in the different directions.  
This idea can be applied in perturbation theory as well (see e.g. 
\cite{karsch89}), and we will follow this method in our analysis
too. It is believed that this procedure ensures a rotationally invariant 
effective action \cite{karsch89,montvay}.

\begin{figure} \begin{center}
\epsfig{file=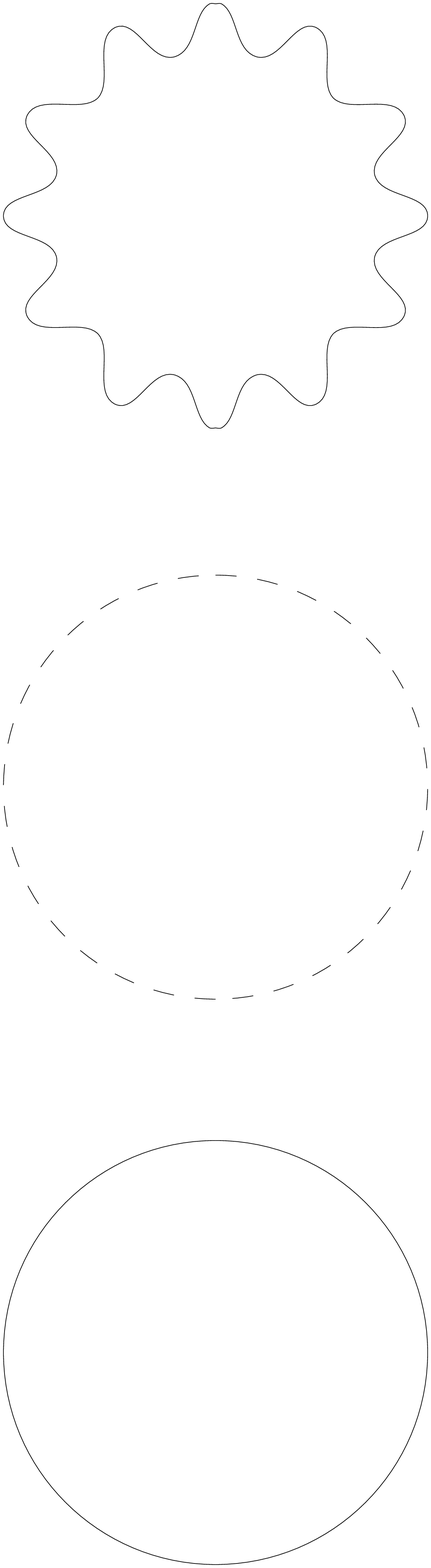,angle=270,width=10.0cm}
\caption{\label{loops}
{\small One-loop diagrams contributing to the effective potential.
The solid line represents the Higgs-, the dashed one
the Goldstone- and the wavy line the vector-boson.
}}
\end{center}\end{figure}

The Feynman-rules of the theory can be derived as usual
(for the three-dimensional case they can be found in
ref. \cite{laine}). In our
analysis we will need the Higgs- and gauge-boson propagators
with one-loop corrections.
The inverse of the tree-level Higgs-boson propagator has the form
\begin{equation}\label{H-propagator}
\Delta_{H,0}(p)^{-1}=m_{H,0}^2+\sum_{i=1}^3 {\hat p}_i^2
+{\gamma_\kappa^2 \over \xi^2}{\hat p}_4^2,
\end{equation}
for the gauge-boson: 
\begin{equation}\label{W-propagator}
\Delta_{W,0,\mu\nu}^{ab}(p)^{-1}
=\delta^{ab}\delta_{\mu\nu}\left[
m_{W,0}^2+\sum_{i=1}^3 {\hat p}_i^2
+{\gamma_\beta^2 \over \xi^2}{\hat p}_4^2
\right].
\end{equation}
Here Feynman-gauge has been used in the Higgs (broken) phase and
\begin{equation}
{\hat p}_i={2 \over a_s}\sin{a_sp_i \over 2},\ \ \ 
{\hat p}_4={2 \over a_t}\sin{a_tp_4 \over 2}.
\end{equation}
The tree-level masses are given by
\begin{equation}
m_{H,0}^2=-{2 \over a_s^2}
\left[{1-2\lambda \over \kappa}\xi-6-2\xi^2\right],\ \ \ 
m_{W,0}^2={m_{H,0}^2\kappa^2 \over 2\lambda\xi\beta}.
\end{equation}
Lattice perturbation theory is conveniently formulated by means of the
parameters $\lambda_c$ and $g$
\begin{equation}
\lambda_c={\lambda\xi \over 4\kappa^2},\ \ \ 
g^2={4 \over \beta}.
\end{equation}

The main goal of the paper is to perform a one-loop analysis of the
theory defined by eq. (\ref{lattice_action}). 
This means first the determination of 
the mass-counterterm. One wants to tune the bare parameters in a way that 
the one-loop renormalized masses are finite in the continuum limit (however,
their values in lattice units vanish $a_sM_{ren}=0$ for
$a_s \rightarrow 0$). At the same time the vacuum expectation value of the 
scalar field will be also zero in lattice units 
($a_s v=0$ for $a \rightarrow 0$) , i.e. we are at the phase transition point
between the spontaneously broken Higgs phase and the SU(2) symmetric phase.
The condition is fulfilled by an appropriate choice of the hopping parameter
(critical hopping parameter). The ratios of the couplings ($\gamma_\beta$
and $\gamma_\kappa$) are still free parameters and can be fixed
by two additional conditions. We demand rotational (Lorenz)
invariance for the scalar and vector propagators on the
one-loop level. This ensures that the propagators with one-loop
corrections have the same form in the $z$ and $t$ directions.
Clearly, arbitrary couplings for different directions
in eq. (\ref{lattice_action}) would not lead to such rotationally
invariant two-point functions.

\begin{figure} \begin{center}
\epsfig{file=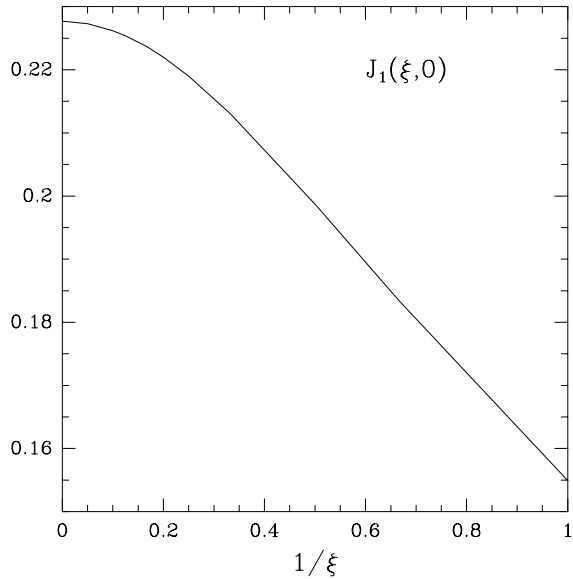,width=8.0cm}
\caption{\label{j_1} 
{\small The lattice integral $J_1(\xi,0)$ (see text) on asymmetric lattices 
as a function of $1/\xi$. 
}}
\end{center}\end{figure}

The most straightforward method to determine the transition point is
the use of the effective potential in the Landau 
gauge. The relevant one-loop graphs are shown in Figure 1.
In this gauge no contribution comes from the ghost fields. The 
direct calculation gives for a constant $\Phi$ field
\begin{equation}
V_{eff}(\Phi)=-{m_{H,0}^2 \over 4}\Phi^2+\lambda_c\Phi^4
+\int_k \left[{1 \over 2} \log({\hat k}^2+12\lambda_c\Phi^2)
+{3 \over 2} \log({\hat k}^2+4\lambda_c\Phi^2)
+{9 \over 2} \log({\hat k}^2+g^2\Phi^2/4)\right],
\end{equation}
where 
\begin{equation}
\int_k \equiv {1 \over (2\pi)^4}
\int_{-\pi/a_s}^{\pi/a_s}d^3k\int_{-\pi/a_t}^{\pi/a_t}dk_4,
\end{equation}
and
\begin{equation}
{\hat k}^2 \equiv \sum_{\mu=1}^4 {\hat k}_\mu^2.
\end{equation}
Note that, in this order, the propagators have no 
$\gamma_{\kappa,\beta}^2/\xi^2$-type corrections in the temporal 
direction. These terms would lead to corrections of higher
order in the critical hopping parameter. The condition
$d^2V_{eff}(\Phi=0)/d\Phi^2=0$ gives the value of the
critical hopping parameter
\begin{equation}\label{critical}
\kappa_c={\xi \over 2(3+\xi^2)}+
{1 \over (3+\xi^2)^2}\left[6\xi J_1(\xi,0)-{\xi^2 \over (3+\xi^2)} \right]
\lambda_c +{9\xi J_1(\xi ,0) \over 16(3+\xi^2)^2}g^2,
\end{equation}
where the notation
\begin{equation}
J_n(\xi,ma_s)=a_s^{4-2n}\int_k {1 \over (m^2+{\hat k}^2)^n}  
\end{equation}
for the dimensionless
lattice integrals on asymmetric lattices is used. For the readers' 
convenience we plot $J_1(\xi,0)$ of eq. (\ref{critical}) in Figure 2 as a
function of $1/\xi$. We have proved that eq. (\ref{critical}) does not 
depend on the choice of the gauge parameter in $R_\xi$ gauge.
For the special case  of symmetric lattice spacings, $\xi=1$, 
our quantum corrections to the critical
hopping parameter reproduce the known result of the isotropic SU(2)-Higgs
model \cite{montvay}. 

\section{Quantum corrections to the anisotropy parameters}

On the one-loop level the propagators of eqs. 
(\ref{H-propagator} and \ref{W-propagator}) receive quantum corrections: 
\begin{equation}\label{propagators}
\Delta_{H,1}(p)^{-1}=\Delta_{H,0}(p)^{-1}+\Sigma_{H,1}(p),\ \ \ 
\Delta_{W,1,\mu\nu}^{ab}(p)^{-1}=
\Delta_{W,0,\mu\nu}^{ab}(p)^{-1}+
\Sigma_{W,1,\mu\nu}^{ab}(p).
\end{equation}
One can demand rotational invariance in the continuum limit, 
$a_s,a_t \rightarrow 0$ at fixed $\xi=a_s/a_t$. The 
corrections to the anisotropies in the kinetic parts of  
eqs. (\ref{H-propagator},\ref{W-propagator}) should be cancelled
by the kinetic parts of the self-energies 
(\ref{propagators}) (cf. \cite{karsch89}). For the Higgs-boson
this can be achieved by demanding  
\begin{equation}\label{condition}
{\gamma_\kappa^2 \over \xi^2}+
{1 \over 2}{\partial^2\Sigma_{H,1}(p) \over \partial p_4^2}\Big|_{p=0}
=1+
{1 \over 2}{\partial^2\Sigma_{H,1}(p) \over \partial p_i^2}\Big|_{p=0},
\end{equation}
where $i=1,2,3$. An analogous condition can be given for the
gauge-boson self-energy as well. These conditions determine the functions
$c_\beta(\xi),c_\kappa(\xi),b_\beta(\xi)$ and $b_\kappa(\xi)$
of eq. (\ref{tree}).

\begin{figure} \begin{center}
\epsfig{file=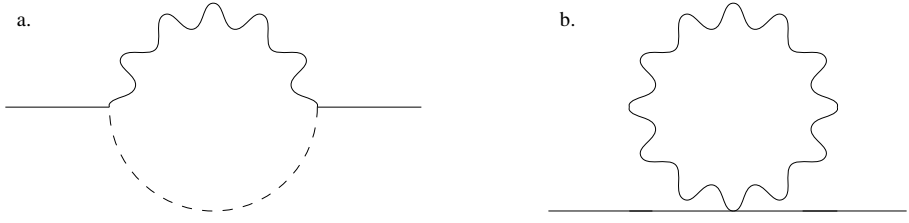,angle=270,width=12.0cm}
\caption{\label{scalar-energy}
{\small Self-energy graphs of the Higgs-boson contributing to the anisotropy
parameters. }}
\end{center}\end{figure}

In this section Feynman-gauge will be used.
The relevant graphs to eq. (\ref{condition}) are shown in Figure 3. 
The diagrams contributing to the analogous condition in the gauge 
sector are given by Figure 4. Note that there are several other 
one-loop contributions to the self-energies; however, those 
graphs give higher order terms in the $g^2,\lambda$ double expansion, or
they are independent of the external momenta.  

The contribution due to the Higgs $\rightarrow$ Goldstone+vector
$\rightarrow$ Higgs graph reads
\begin{equation}\label{graph-a} {3 g^2 \over 4}\int_k
{{\widehat {(p+k)}}^2 \over ({\hat k}^2+M_W^2)({\widehat {(p-k)}}^2+M_W^2)},
\end{equation}
whereas the Higgs $\rightarrow$ vector $\rightarrow$ Higgs
with four-particle vertex gives
\begin{equation}\label{graph-b}
-{3 g^2 \over 4}\int_k
\sum_{\mu=1}^4 \cos^2{p_\mu a_\mu }
{1 \over {\hat k}^2+M_W^2}.
\end{equation}
The contributions to the vector self-energy are much more complicated
(particularly the one due to the vector four-coupling of Figure 4.b);
therefore, they will be not listed here. The corrections to the anisotropies
are obtained as a difference of two integrals --see eq. (\ref{condition})-- 
which is always finite. The functions $c_\beta (\xi)$ and $c_\kappa(\xi)$
of eq. (\ref{tree}) are plotted in Figure 5.

\begin{figure} \begin{center}
\epsfig{file=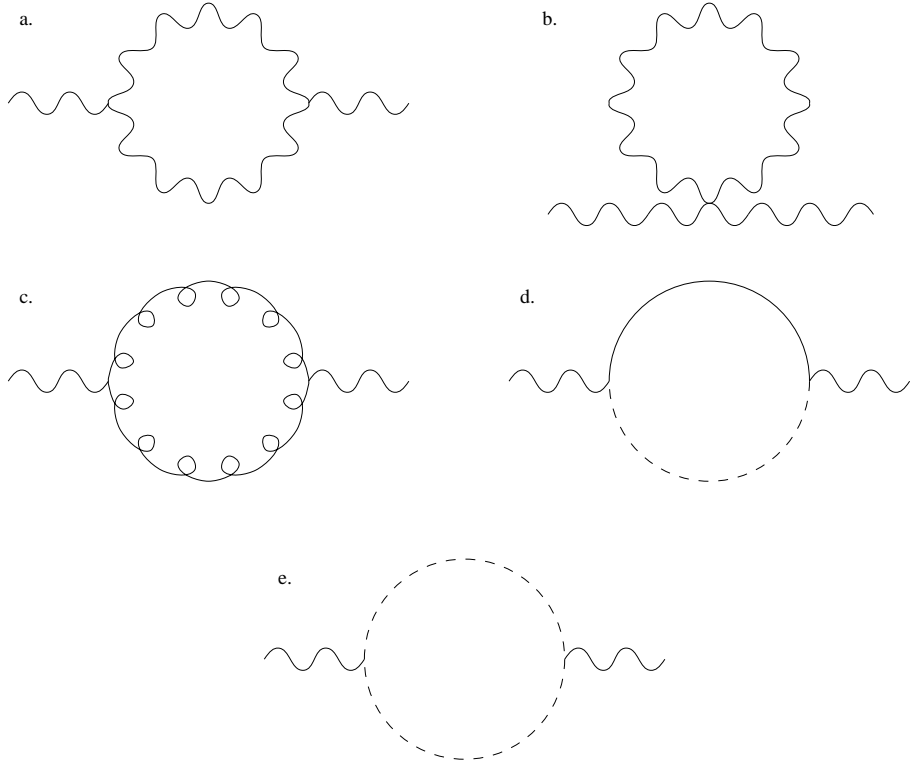,angle=270,width=12.0cm}
\caption{\label{vector-energy}
{\small Self-energy graphs of the vector-boson contributing to the anisotropy
parameters. The curly line represents the ghost field.}}
\end{center}\end{figure}

There are several important features of the result, which
should be mentioned.\hfill\break
{\it a. Masses in the propagators:} a consistent perturbative procedure
on the lattice determines the bare parameters, for which the renormalized 
masses vanish, cf. eq. (\ref{critical}). With these bare couplings other 
quantities, e.g. asymmetry parameters, are determined. However, using
the one-loop renormalized masses ($a_sM_H=a_sM_W=0$) 
in the propagators instead of the bare ones leads to changes in the results,
which are higher order in  $g^2$ and $\lambda$. Therefore, all our
results are given by the integrals with renormalized masses.
\hfill\break
{\it b. $g^2$ and $\lambda$ corrections:} In Figure 5 we have 
given only $c_\beta (\xi)$ and $c_\kappa(\xi)$. The functions
$b_\beta (\xi)$ and $b_\kappa(\xi)$ vanish, thus there are no 
corrections of ${\cal O}(\lambda)$ to the anisotropy parameters.
It is easy to understand this result qualitatively, since only
graphs with two or more scalar self-interaction vertices have non-trivial 
dependence on the external momentum. This feature is connected with
the well-known fact that the $\Phi^4$ theory does not have any 
wave function correction in first
order in the scalar self-coupling. It is worth
mentioning that there is 
only one type of two-loop graph (the setting-sun) which should be 
combined with the one-loop graphs, in order to obtain the whole
${\cal O}(\lambda^2)$ correction.\hfill\break  
{\it c. Pure gauge theory:} The graphs of Figure 4.a, 4.b and 4.c 
are identical with those of the pure gauge theory. Evaluating
these diagrams one reproduces the result of ref. \cite{karsch}
(the function $c_\beta (\xi)$ of the present paper corresponds to
$c_\tau(\xi)-c_\sigma(\xi)$ of ref. \cite{karsch}).  
The most important contribution comes from the self-energy graph with
vector four-coupling of Figure 4.b. Inclusion of the scalar particles
gives only small changes. The relative difference between the $c_\beta (\xi)$
functions for the pure SU(2) theory and for the SU(2)-Higgs model
is typically a few \%.\hfill\break 
{\it d. Quantum corrections to the hopping parameter:} the 
contributions to the hopping parameter come from Figure 3.a and 3.b.
This correction has the same sign and order of magnitude than that of
the gauge anisotropy parameter; however it is somewhat smaller. It is 
possible to combine the anisotropies 
$c_\beta '(\xi)=c_\beta (\xi)-c_\kappa(\xi)$. 
For this choice in the gauge sector and with $\gamma_\kappa =\xi$ the 
rotational invariance can be restored on the one-loop level, choosing the 
appropriate value for the lattice spacing asymmetry $a_s /a_t$. Thus, the 
masses in both directions will be the same. However, the obtained lattice 
spacing asymmetry will then slightly differ from the original $\xi$;
one gets $a_s/a_t= \xi(1-g^2c_\kappa(\xi)/2)+{\cal O}(g^4,\lambda^2)$.


Finally, it should be emphasized that the truly non-perturbative 
analysis of several thermodynamical quantities on lattices with
anisotropic couplings ought to include the non-perturbative 
determination of the anisotropy parameters. This can be
achieved by the study of the correlation functions and/or
static potentials given by Wilson-loops. 

\begin{figure} \begin{center}
\epsfig{file=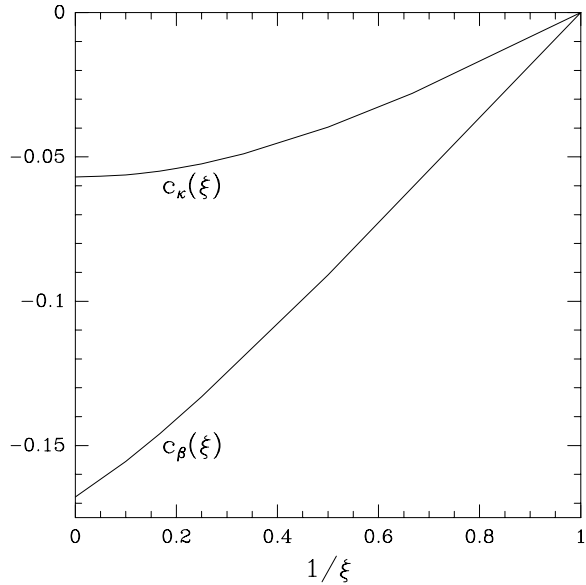,width=8.0cm}
\caption{\label{vector}
{\small $c_\beta (\xi)$ and $c_\kappa(\xi)$ as functions of
$1/\xi$.}}
\end{center}\end{figure}

\section{Discussion}

We have studied the SU(2)-Higgs model on lattices with asymmetric
lattice spacings $a_s \neq a_t$. We have determined the 
${\cal O}(g^2,\lambda)$ corrections to the coupling anisotropy
parameters $\gamma_\beta$ and $\gamma_\kappa$. The corrections
of order $\lambda$ vanish. The $g^2$ corrections 
are quite small. Since the finite temperature electroweak phase transition
is characterized by the dimensionless gauge-coupling $g^2 \approx 0.5$, 
the one-loop perturbative approach to the lattice anisotropy
parameters seems to be satisfactory.

\vspace{.5cm}

Special thanks go to I. Montvay for essential proposals.  
Discussions with K. Kajantie, F. Karsch, A. Patk\'os, M. Shaposhnikov 
and R. Sommer are also acknowledged. This work was partially supported by  
Hungarian Science Foundation grant under Contract  No. OTKA-T016248/7.

\vfill
\end{document}